\documentclass[12pt]{article}
\input{epsf}
\begin{document}

\begin{titlepage}
\begin{flushright}
NSF-KITP-03-38\\
\end{flushright}

\begin{center}
{\Large $ $ \\ $ $ \\
Nonlinear waves in AdS/CFT correspondence}\\
\bigskip\bigskip
{\large Andrei Mikhailov\footnote{e-mail: andrei@kitp.ucsb.edu}}
\\
\bigskip\bigskip
{\it Kavli Institute for Theoretical Physics, University of California\\
Santa Barbara, CA 93106, USA\\
\bigskip
and\\
\bigskip
Institute for Theoretical and 
Experimental Physics, \\
117259, Bol. Cheremushkinskaya, 25, 
Moscow, Russia}\\

\vskip 1cm
\end{center}

\begin{abstract}
We calculate in the strong coupling and large $N$ limit the 
energy emitted by an accelerated external charge
in ${\cal N}=4$ $SU(N)$ Yang-Mills theory, 
using the AdS/CFT correspondence. 
We find that the 
energy is a local functional of the trajectory of the charge. It
coincides up to an overall factor with the Li\'enard
formula of the classical electrodynamics. In the AdS description
the radiated energy is carried by a nonlinear wave on the
string worldsheet for which we find an exact solution.
\end{abstract}
\end{titlepage}

\section{Introduction.}
Wilson loops are very natural observables in gauge
theories.  In the AdS/CFT correspondence they have a clear 
string theoretic interpretation as the boundaries 
of strings in the AdS space 
\cite{ReyYee,MaldacenaWilson,ReyTheisenYee,BCFM,DGO}.
They were used in the recent checks and applications of the AdS/CFT
correspondence 
\cite{ESZ,DG,RSZ,JP,GKP02,Kruczenski,Makeenko,BGK,PZ}.
In our paper we will study the AdS dual of certain timelike
Wilson loops. Timelike Wilson loops correspond to
external sources.
For example, let us consider the ${\cal N}=4$ Yang-Mills
theory on ${\bf R}\times S^3$. 
Let us define  the Wilson loop observable following
\cite{ReyYee,MaldacenaWilson}:
\begin{equation}
W[C]={1\over N}{\rm tr}P\exp i\int_C 
\left(A_{\mu} dx^{\mu}+\Phi_i\theta^i |\dot{x}|\right)dt
\end{equation}
Consider the correlation function 
\begin{equation}
\langle T\left(W[C] {\cal O}(t_1,{\bf n}_1)\;
{\cal O}(t_2, {\bf n}_2),\; \ldots {\cal O}(t_k, {\bf n}_k)
\right)
\rangle
\end{equation}
where $C$ is the contour which runs from $t=-\infty$ to 
$t=+\infty$ at the north pole of the sphere ${\bf n}=(1,0,0,0)$
and returns from $t=+\infty$ to $t=-\infty$ at the south pole
${\bf n}=(-1,0,0,0)$:
\begin{figure}[h]
\begin{center}
\epsffile{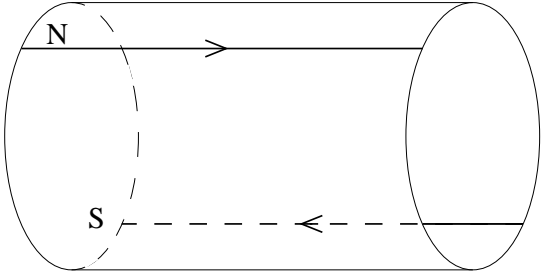}
\caption{Timelike Wilson line.}
\end{center}
\end{figure}
 The insertions ${\cal O}(t_i,{\bf n}_i)$
are some local operators. 
This expression can be interpreted as
a transition amplitude in the Yang-Mills theory on $S^3$
with the external quark and antiquark sitting at the south
and the north pole of the $S^3$. At $t=-\infty$ we have
a ground state of the Yang-Mills on $S^3$ with two sources
which preserves $1\over 2$ of the supersymmetry. 
At $t=t_i$ we act by the operators ${\cal O}(t_i,{\bf n}_i)$
and then we compute the overlap of the resulting state with
the ground state at $t=+\infty$. 

Besides inserting the local operators 
${\cal O}(t_i,{\bf n}_i)$ we can perturb the ground state
by moving the external source. Let us deform the contour $C$
as shown on Fig.2. We leave the component at the south pole unchanged, 
and we create the "wiggle" near the north pole:
\begin{equation}
{\bf n}(t)=
\left( \sqrt{1-\vec{x}^{\;2}(t)},\; x_1(t), x_2(t), x_3(t) \right)
\end{equation}
where $x_1(t), x_2(t), x_3(t)$ are all zero outside
the small interval $-a<t<a$. This corresponds to
the accelerated motion of one of the external charges.
\begin{figure}[h]
\begin{center}
\epsffile{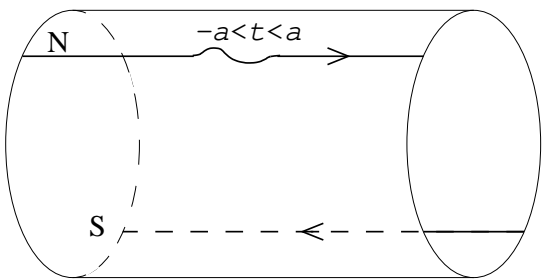}
\caption{Moving one of the sources.}
\end{center}
\end{figure} 
The moving charge
emits radiation, therefore at $t>a$ the system is in some excited state. 
One can try to characterize this excited state and study its 
dependence on the trajectory of the charge.  If this state can be described
quasiclassically one can ask about the energy.
The analogous question in 
classical electrodynamics is
answered by the Li\'enard formula
 \cite{LandauLifshitz,Jackson}:
\begin{eqnarray}\label{Lienard}
&&\Delta E= A
\int_{-\infty}^{\infty} {\ddot{\vec{x}}^{\;2}-
\left.[\dot{\vec{x}}\times \ddot{\vec{x}}]\right.^2\over
(\,1-\dot{\vec{x}}^{\;2}\,)^{\scriptstyle 3\atop  }}dt,\\[5pt]
&&A={2\over 3}e^2\nonumber
\end{eqnarray}
The functional form of this expression follows
from the dimensional analysis and the relativistic invariance
under the assumption that 
the emitted energy is a local functional of the 
particle trajectory. This assumption relies on the fact that 
the Maxwell equations are linear. Indeed, the
field created by the particle at any given moment of its
history is independent of the fields already created earlier;
it just adds to them. Besides that, the electromagnetic field
at any given point $P$ of space depends only on the
trajectory of the particle near the intersection
with the past light cone of $P$; therefore there is no interference
between the electromagnetic waves created at different times. 

The Yang-Mills theory is nonlinear, therefore we might expect
that the energy emitted by the moving source is a 
nonlocal functional of the trajectory. 
At strong coupling we can study this process using 
the AdS/CFT correspondence. The theory with two external
quarks corresponds to the supergravity in $AdS_5\times S^5$
with the  classical string with two boundaries.
At $t<-a$ the worldvolume of the string is $AdS_2\subset AdS_5$
with the Lorentzian signature. The movement of the source
at $t\in [-a,a]$ results in the nonlinear wave
on the string worldsheet. In the large $N$ limit we neglect
the interaction with the closed strings. Therefore all the
energy on the string theory side is  in this nonlinear wave. 
We will find the corresponding exact solution and compute its energy.
 The answer is given
by the Li\'enard formula (\ref{Lienard}) with the overall
coefficient 
\begin{equation}
A={\sqrt{\lambda}\over 2\pi}
\end{equation}
It is surprising that the radiated energy
is a local functional of the trajectory in the strongly
coupled Yang-Mills theory which is highly nonlinear.
This is a manifestation of the integrability of the
classical worldsheet  sigma model \cite{MSW,BPR}.
It would be interesting to see
if the quantum sigma model has an analogous property. 

\vspace{10pt}
\noindent
{\em The plan of the paper.}
In Section 2 we  study the waves on the string worldsheet
in the linearized approximation. In Section 3 we  introduce the
solution for the nonlinear wave and discuss its basic properties.
In Section 4 we  compute the energy and the momentum
carried by the wave. In Section 5 we discuss  
open questions.

\section{Linearized equations.}
\subsection{Two straight lines.}
In this section we will consider the string worldsheet
corresponding to the quark-antiquark pair, and study its
small fluctuations.
$AdS_5$ is the universal covering space of the hyperboloid
\begin{equation}
X_{-1}^2+X_0^2-X_1^2-X_2^2-X_3^2-X_4^2=1
\end{equation}
in ${\bf R}^{2+4}$. We can parametrize this hyperboloid
by $\rho,t,{\bf n}$: $X_{-1}+iX_0=e^{it} \cosh\rho $,
$X_a=n_a\sinh\rho, a=1,2,3,4$. The boundary is at 
$\rho=\infty$, it is a product of $S^1$ parametrized
by $t$ and $S^3$ parametrized by $\bf n$. We will consider
the string worldsheet with two boundaries. One boundary
has ${\bf n}=(0,0,0,1)$ and $t$ going from $-\infty$
to $+\infty$ and the other has ${\bf n}=(0,0,0,-1)$ and
$t$ goes from $+\infty$ to $-\infty$. The corresponding 
worldsheet is given by the equation:
\begin{equation}
X_1=X_2=X_3=0
\end{equation}
\begin{figure}
\begin{center}
\epsffile{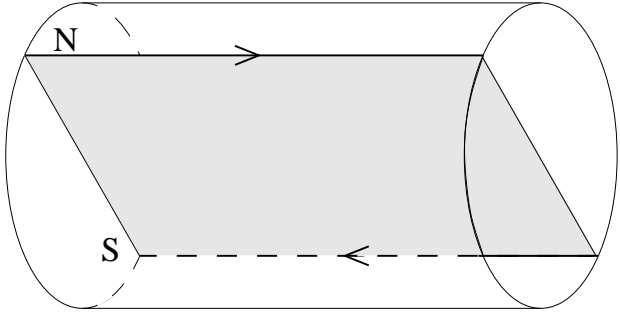}
\caption{$AdS_2\subset AdS_5$}
\end{center}
\end{figure}
It may be parametrized by $X_{-1}$, $X_0$ and $X_1$
satisfying $X_{-1}^2+X_0^2-X_1^2=1$. The induced metric
makes it $AdS_{1+1}$. The Poincare coordinates
cover a wedge of the AdS space:
\begin{eqnarray}\label{PoincareCoordinates}
&v=&(X_{-1},X_0,X_1,X_2,X_3,X_4)=\\[5pt]
&&=\left[ {1-x_{\mu}^2+h^2\over 2h}, {x_0\over h};
{\vec{x}\over h}, {1+x_{\mu}^2-h^2\over 2h}\right]
\end{eqnarray}
The trajectory of the quark on the south pole is not covered
at all by these coordinates. A part of the trajectory
of the northern quark is covered, and is given by the equations:
$x_1=x_2=x_3=0$. The trajectory is a straight line.
The string worldsheet is the upper half-plane ending on this
straight line.
\subsection{Small fluctuations.}
The Poincare coordinates are convenient for studying the 
linearized wave equations on the worldsheet.
We will describe the wave by the 
displacement $x_i(x_0,h)$. The variation of the area functional is:
\begin{equation}
S=\int {dx_0 dh\over h^2}\left[
 \left( {\partial x_i\over\partial x_0}\right)^2
-\left( {\partial x_i\over\partial h}\right)^2 \right]
\end{equation}
The wave equation for $x_i$ is:
\begin{equation}
{\partial^2 x_i\over \partial x_0^2}
-{\partial^2 x_i\over\partial h^2}+
{2\over h}{\partial x_i\over\partial h}=0
\end{equation}
It follows that $y_i={1\over h}{\partial x_i\over\partial h}$
is a harmonic function: $y_i=-x_{i+}''(x_0+h)-x_{i-}''(x_0-h)$.
Therefore
\begin{equation}\label{LinearizedSolution}
x_i(x_0,h)=-hx_{i+}'(x_0+h)+hx_{i-}'(x_0-h)+
x_{i+}(x_0+h)+x_{i-}(x_0-h)
\end{equation}
Notice that $x_i(x_0,0)=x_{i+}(x_0)+x_{i-}(x_0)$
and 
$\left.{\partial^3\over\partial h^3}\right|_{h=0}
x_i(x_0,h)=-2x_+'''(x_0)+2x_-'''(x_0)$. Therefore
the shape of the minimal surface is uniquely determined
by the shape of the boundary and the third derivative at the
boundary (see also Section 6 of \cite{sleg}).

The solution
(\ref{LinearizedSolution}) can be thought of as a sum
of the left-moving wave parametrized by $x_{i+}$
and the right-moving wave parametrized by $x_{i-}$. Suppose
that the quark was at rest for the times $x_0<0$, 
then moved a little bit at $0<x_0<a$ and then returned to its 
original position and stayed at rest for $x_0>a$. 
Such a motion will create a right-moving
wave, and the left-moving part will be zero (that is, $x_{i+}=0$).
The fluctuation of the worldsheet will be localized in the
strip $x_0-h\in [0,a]$. If we stay in the Poincare patch
we would see only a part of the story. Let us follow the
wave beyond the Poincare patch.  
The hyperboloid is covered by the two copies of the Poincare
patches which we will denote ${\cal P}_0$ and ${\cal P}_1$.
In the parametrization (\ref{PoincareCoordinates})
${\cal P}_0$ covers the region $h>0$ 
and ${\cal P}_1$ covers $h<0$. The wave is localized in a strip
$x_0-h\in [0,a]$. In this strip the coordinates 
$u=h^{-1}$ and $v=x_0-h$ are nonsingular. We can continue the solution
beyond the Poincare patch by relaxing the condition $u>0$ 
and considering
arbitrary $u$. 
\begin{figure}
\begin{center}
\epsffile{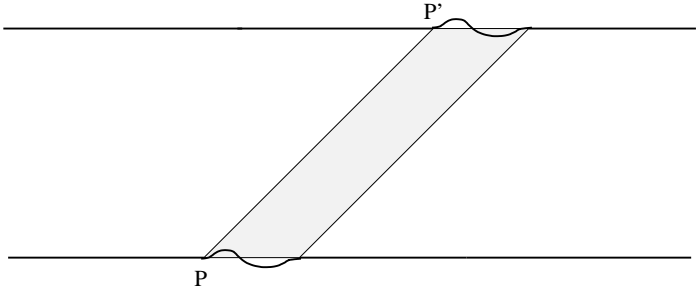}
\caption{A right-moving wave.}
\end{center}
\end{figure}
This gives a left-moving wave in the second
Poincare patch ${\cal P}_1$. 
The full solution in the hyperboloid is:
\begin{equation}\label{Global}
\begin{array}{l}
x_i(x_0,h)=
hx'_i(x_0-h)+x_i(x_0-h) \;\;\; {\rm in}\;\;\;{\cal P}_0 \\[5pt]
x_i(x_0,h)=
-hx'_i(x_0+h)+x_i(x_0+h)\;\;\; {\rm in}\;\;\;{\cal P}_1
\end{array}
\end{equation}
The boundary condition in ${\cal P}_1$ is the same as 
it was in ${\cal P}_0$. The solution (\ref{Global})
has the following physical meaning. We create a wave on
the string worldsheet by the small motion $\delta x_i(x_0)$ 
of one of the boundaries
at time $x_0=0$. The wave propagates forward in time
and reaches the second boundary where it is totally absorbed
by the identical motion $\delta x_i(x_0)$ 
of the second boundary (see Figure 4).

In order to better understand this solution it is useful to
remember a simple geometric feature of the boundary
${\bf R}\times S^3$. The geometry
of ${\bf R}\times S^3$ and the propagation of the
massless fields on ${\bf R}\times S^3$ was considered in
great detail in \cite{LM}. Given a point 
$P\in {\bf R}\times S^3$ let us consider the future light cone of 
$P$. Because of the symmetries of $S^3$, the light rays
emitted at point $P$ will refocus and converge at some other
point $P'\in {\bf R}\times S^3$. The correspondence
$P\mapsto P'$ is a symmetry of ${\bf R}\times S^3$, which
is denoted $Z$ in \cite{LM}. It is a combination of the
 reflection  ${\bf n}\to -{\bf n}$ of $S^3$ and the shift
$t\to t+\pi$ in ${\bf R}$ (we choose the radius of $S^3$
to be $1$). In these notations our wave is generated by
the small perturbation $\delta x_i$ of the boundary at the
point $P$ with the coordinates 
$x_{\mu}=0$, 
propagates with the speed of light forward in time and is
then totally absorbed by the identical perturbation $\delta x_i$ 
of the second boundary at the point $P'=Z.P$. 
If we do not perturb the second boundary at $P'$ then
the wave will not get absorbed. Instead it will reflect
from the second boundary and travel back towards to the
first boundary.  To describe the corresponding solution
we have to consider the right moving wave generated by the
perturbation $-\delta x_i$ at the point $P'$, and
add it to the solution (\ref{Global}). This solution 
corresponds to the wave created by the perturbation
$\delta x_i$ at $P$, then traveling with the speed of
light towards the second boundary, then being reflected
from the second boundary near the point $P'=Z.P$, then
coming back to the first boundary and being cancelled there
by the perturbation $-\delta x_i$ at the point 
$P''=Z^2.P$. Of course, we could also add a right moving
wave generated by $\delta x_i$ at $P''$, and get a reflected
wave again. 

Let us summarize our discussion of the linear waves.
The small perturbation $\delta x_i$ 
of the boundary at the point
$P$ will create a wave moving forward in time with the
speed of light, bouncing back and forth between the
two boundaries. We can absorb this wave
by the perturbation $\delta x_i$ at some 
point $Z^{2n+1}.P$, or by the perturbation $-\delta x_i$
at some point $Z^{2n}.P$ --- see Figure 5.
\begin{figure}
\begin{center}
\epsffile{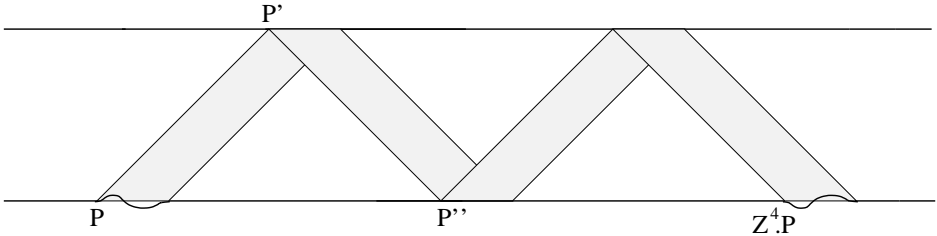}
\caption{A linear wave.}
\end{center}
\end{figure}

\section{Nonlinear wave.}
In this section we will construct the extremal surface
corresponding to the solution (\ref{Global}) for a finite
perturbation. It is a nonlinear wave on
the worldsheet emitted at some point $P$ of the boundary
and absorbed at the point $Z.P$.
\vspace{10pt}

\noindent
{\em The construction.}
Let us realize the AdS space as a hyperboloid in ${\bf R}^{2+4}$
consisting of the vectors $v\in {\bf R}^{2+4}$
satisfying $||v||^2=1$. The AdS space is actually a universal
covering space of this hyperboloid. But 
our solution all fits in a simply-connected domain of 
the hyperboloid and therefore it is not important for our purposes
to distinguish between the hyperboloid and its universal
covering.
The boundary consist of the light rays in ${\bf R}^{2+4}$. 
Suppose that $l(\tau)$ is a one-parameter 
family of the lightlike vectors representing the boundary
of the string worldsheet. We will choose the parameter $\tau$
so that $\left({d l\over d\tau}, {dl\over d\tau}\right)=1$.
Consider the following
two-dimensional surface parametrized by the two real
parameters $\tau$ and $\sigma$:
\begin{equation}\label{Surface}
v(\tau,\sigma)=-\dot{l}(\tau)+\sigma l(\tau)
\end{equation}
Let us prove that this surface is extremal. Indeed, let us
start with computing the induced metric:
\begin{equation}
ds^2=||dv(\tau,\sigma)||^2=(||\ddot{l}||^2+\sigma^2)d\tau^2
+2 d\tau d\sigma
\end{equation}
The area is therefore just $\int d\tau d\sigma$. 
We have to prove that the area does not change under
the infinitesimal variations of the surface. We can 
describe the infinitesimal variations by the normal
vector field $\xi(\tau,\sigma)$. We should have 
$(\xi, -\dot{l}+\sigma l)=0$ (because $\xi$ is tangent
to the hyperboloid) and also $(\xi, l)=0$ and
$(\xi, -\ddot{l}+\sigma\dot{l})=0$ (because $\xi$
is normal to the surface). It follows that 
$(\xi,l)=(\xi,\dot{l})=(\xi,\ddot{l})=0$. 
The metric on the perturbed surface is,
to the first order in $\xi$:
\begin{equation}
ds^2=||dv(\tau,\sigma)||^2=
\left(||\ddot{l}||^2+\sigma^2+
2\left(\xi, {d^3l\over d\tau^3}\right)\right)d\tau^2
+2 d\tau d\sigma +o(\xi)
\end{equation}
Therefore the variation of the area functional is zero.

The regularized area of this surface is zero. 
Indeed, let us regularize the area functional
by restricting the $\sigma$ coordinate to run
from $-S$ to $S$. According to the prescription
in  \cite{ReyYee,MaldacenaWilson,DGO} the regularized area
is the area minus the length of the boundary.
The area is $2S\int d\tau$ and the
length of the boundary is 
$2\int d\tau \sqrt{S^2+||\ddot{l}||^2 }$. 
Therefore the area is cancelled by the length
of the boundary in the limit when $S$ is large.

Let us rewrite (\ref{Surface}) in the Poincare coordinates:
\begin{equation}\label{RightMoving}
\begin{array}{l}
h(\tau,\sigma)=\sigma^{-1}\\[5pt]
x_{\mu}(\tau,\sigma)=\sigma^{-1}\dot{x}_{\mu}(\tau)
+x_{\mu}(\tau)
\end{array}
\end{equation}
Here we have assumed that $\dot{x}_{\mu}\dot{x}^{\mu}=1$.
This is a right moving wave. The left moving wave is
\begin{equation}\label{LeftMoving}
\begin{array}{l}
h(\tau,\sigma)=\sigma^{-1}\\[5pt]
x_{\mu}(\tau,\sigma)=-\sigma^{-1}\dot{x}_{\mu}(\tau)
+x_{\mu}(\tau)
\end{array}
\end{equation}
The left moving wave in the global coordinates: 
$v(\tau,\sigma)=\dot{l}(\tau)+\sigma l(\tau)$.
\vspace{10pt}

\noindent
{\em The solution may have wedges.}
Suppose that the second derivative $\ddot{x}$ of the
boundary contour is not continuous. Then the resulting
wave is not smooth.
The light ray originating from the point where $\ddot{x}$ has a jump
is a "wedge" on the worldsheet; the first derivative of the
embedding jumps across the wedge. In general, the Lorentzian
extremal surfaces may have wedges. The wedge
does not have to be a null-geodesic in AdS, but it has to be a 
lightlike curve. Indeed, let us consider a surface 
composed of two smooth extremal surfaces glued along the
lightlike curve. We claim that this surface is extremal.
Let us introduce on both surfaces the coordinates $x_+, x_-$
so that  they agree on the intersection and the induced metric
is $f(x_+,x_-)dx_+dx_-$. Consider the variation of the
area under the infinitesimal deformation 
$\delta X^{\mu}(x_+,x_-)=\xi^{\mu}(x_+,x_-)$. Because each of the two
surfaces are extremal, the possible change in the area comes from
the boundary term which has a form
$$
\int_L dx_+ \xi_{\mu}{\partial\over\partial x_+}X^{\mu}
$$
where the integral is over the intersection. But
$X^{\mu}$ is continuous across the intersection, therefore
${\partial\over\partial x_+}X^{\mu}$ is also continuous.
Therefore the boundary terms from for the two surfaces
cancel each other.

\vspace{10pt}

\noindent
{\em A change in boundary conditions.}
The nonlinear wave which we have constructed propagates from one
boundary to the other. Both boundaries are perturbed.
What happens if we perturb only one boundary, leaving the other
one a straight line? Infinitesimal perturbation will create a
wave localized in a strip, which bounces forever back and forth
between the boundaries --- see Figure 5.
What can we say about the solution of the nonlinear equations?
 We  know that a part of the worldsheet has
the same shape as the right-moving wave.
\begin{figure}
\begin{center}
\epsffile{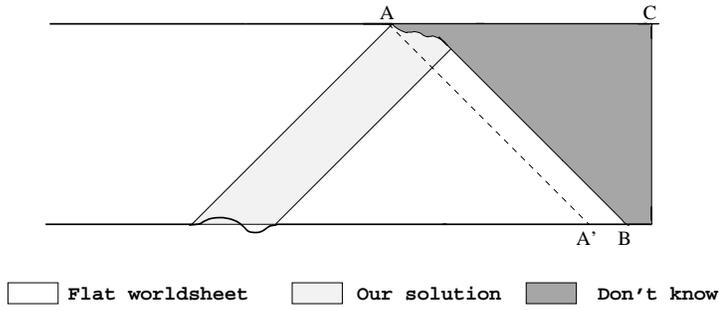}
\caption{Changing the boundary conditions on the upper boundary.}
\end{center}
\end{figure}
The solution is scetched on Figure 6. 
The curve $AB$ is a light ray. We know the solution
on the left of $AB$. Indeed, the boundary condition
on the upper boundary is different from the boundary 
condition for our solution (\ref{Surface}) only on the halfline
$AC$. The difference can only affect the solution in the
future of this halfline, which is the dark area of Figure 6.

\begin{figure}
\begin{center}
\epsffile{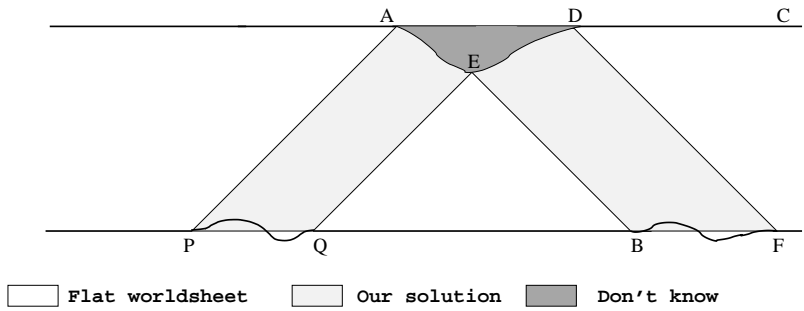}
\caption{Reflection of nonlinear wave.}
\end{center}
\end{figure}
There is an additional qualitative argument which implies
that most of the solution in the dark area is actually 
a left moving wave, see Figure 7.
The boundary conditions at the upper boundary leave undefined
the third derivative $\alpha^i={\partial^3 x^i\over\partial h^3}$
(see for example \cite{sleg}). To the left of the point $A$
the worldsheet is flat. Therefore we should take $\alpha^i=0$
to the left of $A$.
 To the right
of A let us  adjust this third
derivative in such a way that  the resulting minimal
surface contains the interval $AE$ of the light ray.
The resulting surface will also contain an interval $ED$
of some other light ray. Now, let us continue the solution
below the null curve $AED$ as shown on Figure 7: three flat regions
and two waves, one right-moving and one left-moving.
We do not have a problem adding the left-moving nonlinear wave
and the right-moving nonlinear wave because they do not overlap.
 The only "dark" region where we do not know the solution
is the triangle $AED$. We need to know the solution in this
region if we want to learn the relation between the 
trajectory of the source between points $P$ and $Q$ and
the trajectory between the points $B$ and $F$.

\vspace{10pt}
\noindent
{\em Scattering.}
Similar arguments can be applied to the scattering of the
right moving wave on the left moving wave --- see Figure 8.
The left moving wave collides with the right moving wave at
some point $B$. We know the solution for the nonlinear
wave, therefore we know the shape of the null intervals
$BA$ and $BC$. In principle one can find using the methods of
\cite{FT} the shape of the worldsheet inside the diamond $ABCD$.
The intervals $AB$, $BC$, $CD$ and $DA$ are all null curves.
One should then adjust the trajectory of the source 
in the intervals $XY$ and $ZW$ so that the left- and right-moving
waves created by the source contain null intervals $AD$ and $DC$.
\begin{figure}
\begin{center}
\epsffile{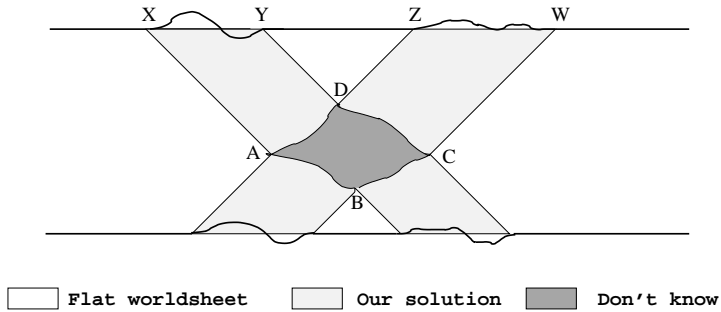}
\caption{Scattering}
\end{center}
\end{figure}
It may happen that the light rays $YD$ and $ZD$ do not belong
to the same vertical ($\vec{x}=0$ in Poincare coordinates) plane.
In this case the interval $YZ$ has to be a part of a hyperbola,
rather than a straight line.

When the left moving wave passes through the right moving wave,
the shapes of both waves are distorted. 
A simple way to see it is to look at
the propagation of the left moving light rays in the background
of the right moving nonlinear wave. One can see that the image 
of the boundary is distorted after
the light rays pass through the right-moving
nonlinear wave. This means that even a linear left-moving wave
is generally distorted by the nonlinear wave.
\section{Energy of the wave.}
We will compute the energy of the wave in the Poincare
coordinates, with respect to the time $x^0$ defined
in the Poincare patch.
Let us parametrize the worldsheet by $T=x_0$ and $h$.
The energy is
\begin{equation}\label{GeneralEnergy}
E={\sqrt{\lambda}\over 2\pi}\int_0^{\infty} {dh\over h^2} 
{1+\left({\partial \vec{x}\over\partial h}\right)^2
\over \sqrt{1-\left({\partial \vec{x}\over\partial T}\right)^2
+\left({\partial\vec{x}\over\partial h}\right)^2
-\left({\partial\vec{x}\over\partial T}\right)^2
\left({\partial\vec{x}\over\partial h}\right)^2
+\left({\partial\vec{x}\over\partial T}\cdot
{\partial\vec{x}\over\partial h}\right)^2}}
\end{equation}
To compute this integral on our solution it is useful
to change the coordinates from $(T,h)$ to $(T,t)$ (see Figure 9):
\begin{figure}
\begin{center}
\epsffile{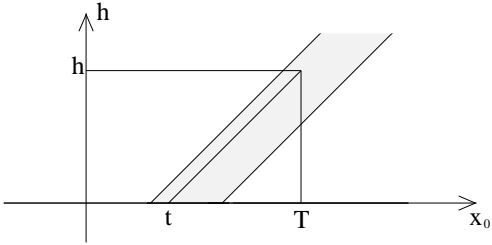}
\caption{Notations for the calculation of the energy.}
\end{center}
\end{figure}
\begin{equation}
h(T,t)=(T-t)\sqrt{1-\left({d\vec{x}(t)\over dt}\right)^2}
\end{equation}
where $\vec{x}(t)$ is the trajectory of the endpoint of the string.
In these coordinates
\begin{equation}
\vec{x}(T,t)=\vec{x}(t)+(T-t){d\vec{x}(t)\over dt}
\end{equation}
A straightforward calculation gives:
\begin{eqnarray}
&&\left.{\partial \vec{x}\over\partial T}\right|_h=
\vec{v}+(T-t){1-\vec{v}^2\over 
1-\vec{v}^2+(T-t)(\vec{v}\cdot\vec{a})}\;\vec{a}\\[5pt]
&&\left.{\partial \vec{x}\over\partial h}\right|_T=
-(T-t)
{\sqrt{1-\vec{v}^2}\over 1-\vec{v}^2+(T-t)(\vec{v}\cdot \vec{a})}
\vec{a}
\end{eqnarray}
Substituting $\left.{\partial \vec{x}\over\partial T}\right|_h$
and $\left.{\partial \vec{x}\over\partial h}\right|_T$
into (\ref{GeneralEnergy}) and discarding a total derivative 
we get:
\begin{equation}\label{Energy}
E={\sqrt{\lambda}\over 2\pi}\int dt {\ddot{\vec{x}}^{\;2}-
\left.[\dot{\vec{x}}\times \ddot{\vec{x}}]\right.^2\over
(\,1-\dot{\vec{x}}^{\;2}\,)^{\scriptstyle 3\atop  }}
\end{equation}
This expression has the same form 
as the Li\'enard formula for the energy emitted by the moving charge 
in the classical electrodynamics 
\cite{LandauLifshitz,Jackson}.
One can also compute the momentum:
\begin{eqnarray}
&&\vec{P}={\sqrt{\lambda}\over 2\pi}
\int_0^{\infty}{dh\over h^2}
{{\partial\vec{x}\over\partial T} +
\left({\partial\vec{x}\over\partial h}\right)^2
{\partial\vec{x}\over\partial T}-
\left({\partial\vec{x}\over\partial T}\cdot
{\partial\vec{x}\over\partial h}\right)
{\partial\vec{x}\over\partial h}
\over \sqrt{1-\left({\partial \vec{x}\over\partial T}\right)^2
+\left({\partial\vec{x}\over\partial h}\right)^2
-\left({\partial\vec{x}\over\partial T}\right)^2
\left({\partial\vec{x}\over\partial h}\right)^2
+\left({\partial\vec{x}\over\partial T}\cdot
{\partial\vec{x}\over\partial h}\right)^2}}
\\[5pt]
&&={\sqrt{\lambda}\over 2\pi}\int dt {\ddot{\vec{x}}^{\;2}-
\left.[\dot{\vec{x}}\times \ddot{\vec{x}}]\right.^2\over
(\,1-\dot{\vec{x}}^{\;2}\,)^{\scriptstyle 3\atop  }}
\dot{\vec{x}}
\end{eqnarray}
This expression actually follows from (\ref{Energy})
and the Lorentz invariance. The relativistic formula is
$
P^{\mu}={\sqrt{\lambda}\over 2\pi}\int 
{d^2 x_{\sigma}\over ds^2}{d^2 x^{\sigma}\over ds^2}
dx^{\mu}
$
where $s$ is a proper time.

\section{Discussion.}
We have studied the nonlinear waves which are excitations
of the extremal surface $AdS_{2}\subset AdS_{5}$.
In some sense all the extremal surfaces in $AdS_{5}$
which are sufficiently close to $AdS_{2}$ can be obtained
as "nonlinear superpositions" of these waves.
Is it possible to give a similar description
of the extremal surfaces extending in $S^5$?
Perhaps they can be obtained as excitations
of the supersymmetric extremal surfaces introduced
in \cite{Zarembo}.

It would be very interesting to understand the
role of these nonlinear waves in the 
quantum  worldsheet theory \cite{BPR}. If the quantum CFT is integrable,
then one should probably expect that the radiated energy is a local
functional of the trajectory in the large $N$ theory at finite $\lambda$.
This raises the possibility of a check by a perturbative field theory 
computation. 
Actually at finite  $\lambda$ one cannot literally ask the question
about the radiated energy. The emitted radiation will be in some
quantum state which does not necessarily have a definite energy.
But one can ask for example about the total probability of emitting
a quant of radiation.

We hope to return to these questions in a future publication.

\section*{Acknowledgements.}
I want to thank H.~Friedrich, D.J.~Gross and 
J.~Polchinski for very interesting discussions. This
research was supported in part by the National Science Foundation under
Grant No. PHY99-07949 and in part
by the RFBR Grant No. 00-02-116477 and in part by the 
Russian Grant for the support of the scientific schools
No. 00-15-96557.

\end{document}